\documentclass[final,5p,times,twocolumn]{elsarticle}

\usepackage{amssymb}
 \usepackage{amsmath}
\usepackage[ansinew]{inputenc}
\usepackage[T1]{fontenc}
\usepackage[range-units = single,separate-uncertainty = true]{siunitx}
\usepackage{subcaption}
\usepackage{makecell}
\usepackage{cite}
\usepackage{graphicx}
\journal{Nuclear Physics B}

\begin{document}

\begin{frontmatter}



\title{Performance of the most recent Microchannel-Plate PMTs for the PANDA DIRC detectors at FAIR}


\author[1]{S.~Krauss}
\author[1]{M.~B\"ohm}
\author[1]{K.~Gumbert}
\author[1]{A.~Lehmann}
\author[1]{D.~Miehling}
\author[2]{A.~Belias}
\author[2]{R.~Dzhygadlo}
\author[2]{A.~Gerhardt}
\author[2]{D.~Lehmann}
\author[2,3]{K.~Peters}
\author[2]{G.~Schepers}
\author[2]{C.~Schwarz}
\author[2]{J.~Schwiening}
\author[2]{M.~Traxler}
\author[2,3]{Y.~Wolf}
\author[4]{L.~Schmitt}
\author[5]{M.~D\"uren}
\author[5]{A.~Hayrapetyan}
\author[5]{I.~K\"oseglu}
\author[5]{M.~Schmidt}
\author[5]{T.~Wasem}
\author[6]{C.~Sfienti}
\author[7]{A.~Ali}

\affiliation[1]{organization={Friedrich Alexander-University of Erlangen-Nuremberg, Erlangen, Germany}}
\affiliation[2]{organization={GSI Helmholtzzentrum f\"ur Schwerionenforschung GmbH, Darmstadt, Germany}}
\affiliation[3]{organization={Goethe-University, Frankfurt, Germany}}
\affiliation[4]{organization={FAIR, Facility for Antiproton and Ion Research in Europe, Darmstadt, Germany}}
\affiliation[5]{organization={II. Physikalisches Institut, Justus Liebig-University of Giessen, Giessen, Germany}}
\affiliation[6]{organization={Institut f\"ur Kernphysik, Johannes Gutenberg-University of Mainz, Mainz, Germany}}
\affiliation[7]{organization={Helmholtz-Institut Mainz, Germany}}


\begin{abstract}
	
	In the PANDA experiment at the FAIR facility at GSI two DIRC (Detection of Internally Reflected Cherenkov light) detectors will be used for $\mathrm{\pi/K}$ separation up to \SI{4}{\giga \eV \per c}. Due to their location in a high magnetic field and other stringent requirements like high detection efficiency, low dark count rate, radiation hardness, long lifetime and good timing, MCP-PMTs (microchannel-plate photomultiplier) were the best choice of photon sensors for the DIRC detectors in the PANDA experiment. This paper will present the performance of some of the latest 2x\SI{2}{inch \squared} MCP-PMTs from Photek and Photonis, including the first mass production tubes for the PANDA Barrel DIRC from Photonis. Performance parameters like the collection efficiency (CE), quantum efficiency (QE), and gain homogeneity were determined. The effect of magnetic fields on some properties like gain and charge cloud width was investigated as well. Apart from that the spatial distribution of many internal parameters like time resolution, dark count rate, afterpulse ratio, charge sharing crosstalk and recoil electrons were measured simultaneously with a multihit capable DAQ system. The latest generation of Photonis MCP-PMTs shows a unexpected "escalation" effect where the MCP-PMT itself produces photons.
\end{abstract}

%


\end{frontmatter}


\section{Introduction} \label{ch:1}
	
	PANDA is one of the main experiments at the FAIR facility at GSI and will study different aspects in QCD using an antiproton beam in the momentum range of \SIrange{1.5}{15}{\giga\eV\per c} colliding with a stationary target. The PANDA detector consists of a target spectrometer and a forward spectrometer. Two Cherenkov detectors of the DIRC type \citep{coyle}, a cylindrically shaped Barrel DIRC \citep{roman2022} around the interaction region and an Endcap Disc DIRC (EDD) \citep{ilknur2020} covering the forward hemisphere, will be used for particle identification, in particular for the separation of $\mathrm{\pi/K}$ up to \SI{4}{\giga \eV \per c} \citep{pandatpr, pandappr}.
	Due to some stringent boundary conditions, and since the focal planes of both DIRC detectors reside in a \SI{\sim 1}{\tesla} magnetic field, microchannel-plate photomultipliers (MCP-PMTs) are the only viable photo detector candidates. After the long-standing aging issue was overcome \citep{LEHMANN2017570} by coating the MCPs with an ultra-thin layer of alumina and/or magnesia by applying an ALD (atomic layer deposition) technique \citep{BEAULIEU200981} to the MCP pores the performance of recent devices was additionally optimized in other important parameters like, e.g., the collection efficiency and the QE and gain homogeneity across the active surface. This and other advantageous properties in terms of radiation hardness, an excellent time resolution, a low dark count rate and in particular their favorable gain behavior inside magnetic fields make MCP-PMTs most suitable for the PANDA DIRCs \citep{BADtdr, EDDTDR}.

\section{Measurements of various parameters and their results} \label{ch:2}
	
	The following sections describe the measurement procedures and the results for the most recent 2x\SI{2}{inch \squared} MCP-PMTs with an anode configuration of 8x8 pixels from Photek and Photonis. 

	\subsection{Lifetime, quantum efficiency and gain} \label{ch:2.1}
		
		In the lifetime measurements the MCP-PMTs are under continous illumination over a long period of time with a single photon rate of \SI{\sim 1}{\mega \Hz \per \cm \squared} at a wavelength of \SI{460}{\nano \m} while monitoring the integrated anode charge (IAC) and measuring the spectral and spatial QE distribution in regular time intervals. In 10 years of PANDA operation time a MCP-PMT has to withstand \SI{\geq 5}{\coulomb \per \cm\squared} IAC without suffering any photocathode damage.
		
		\begin{figure}[ht]
			\centering
			\includegraphics[width=\linewidth]{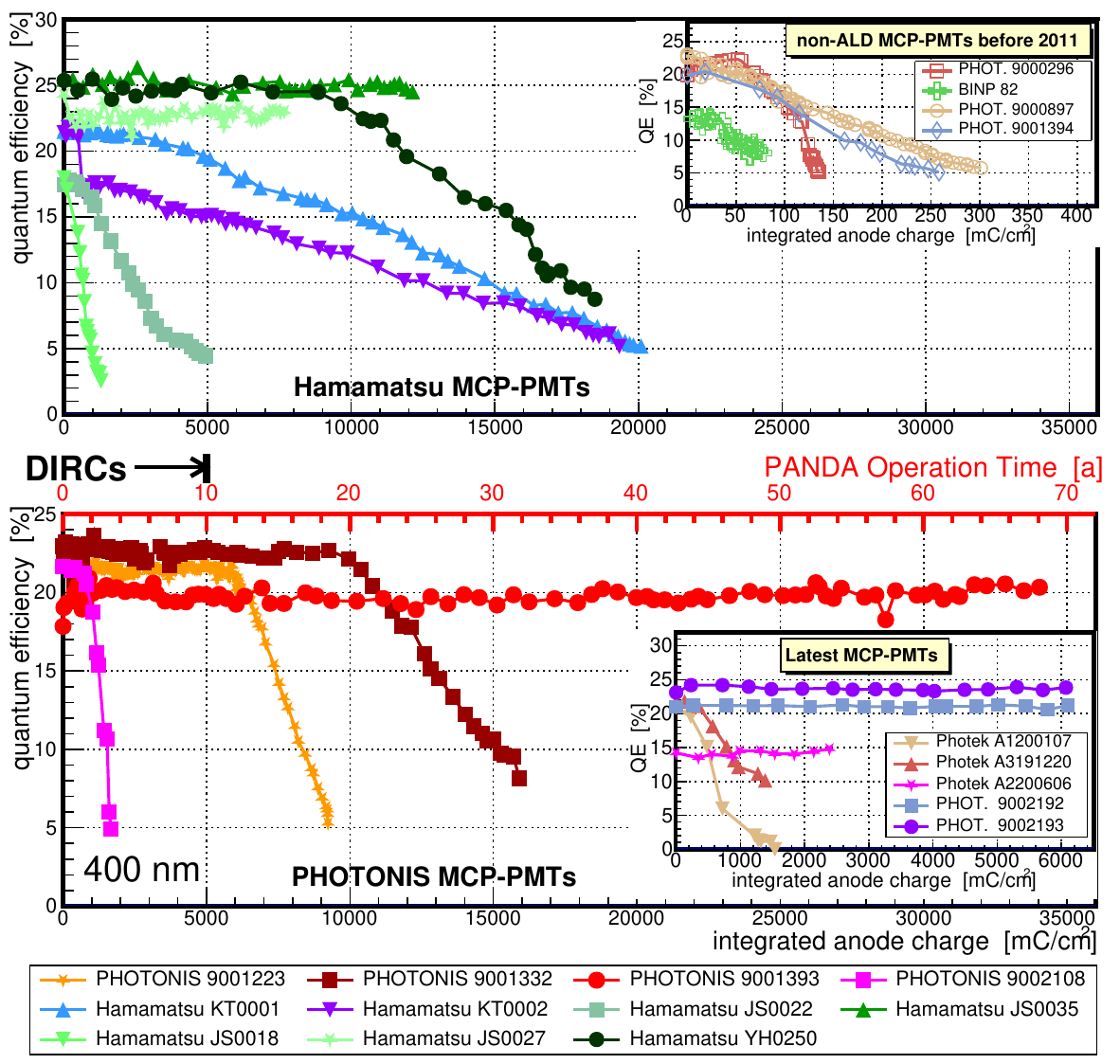} 
			\caption{QE (400 nm) versus IAC; lifetime data for various ALD-coated MCP-PMTs from different vendors compared to non-ALD PMTs (upper right).}
			\label{fig:lifetime}
		\end{figure}	
		
		Figure \ref{fig:lifetime} shows the current status of the lifetime performance for various MCP-PMTs from different vendors. When comparing non-ALD tubes to tubes with ALD coating a tremendous increase of collected IAC without QE loss is visible (compare also to Refs. \citep{Kishimoto:2006mg, Britting:2011zz, Uhlig:2012rk, 1748-0221-11-05-C05009}). The by far best-performing MCP-PMT is Photonis 9001393, which has collected almost \SI{35}{\coulomb \per \cm\squared} IAC which would correspond to nearly 70 years of PANDA operation time. Both latest Photonis tubes 9002192 \& 9002193 surpassed the requirement without any QE decrease as well. Also seen in fig. \ref{fig:lifetime} the Photek A1200107 and A3191220 tubes show a continous QE decrease from the beginning of illumination.
		
		\begin{figure}[ht]
			\centering
			\subcaptionbox{Before illumination. \label{fig:qeA31}}[0.49\linewidth]{\includegraphics[width=\linewidth]{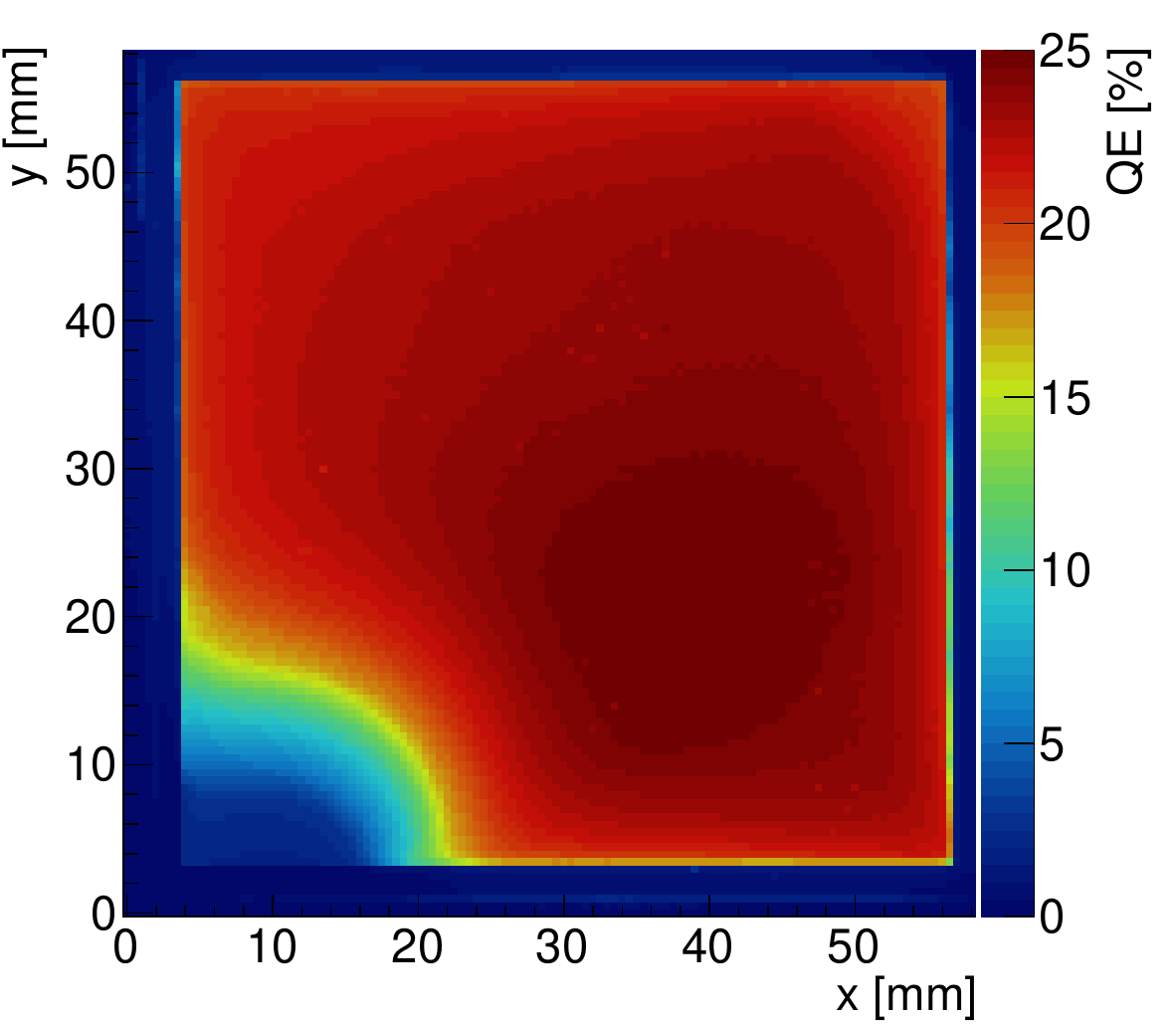}} %
			\subcaptionbox{Right half illuminated with \SI{1.4}{\coulomb \per \cm\squared} IAC, left half was permanently masked. \label{fig:qeA32}}[0.49\linewidth]{\includegraphics[width=\linewidth]{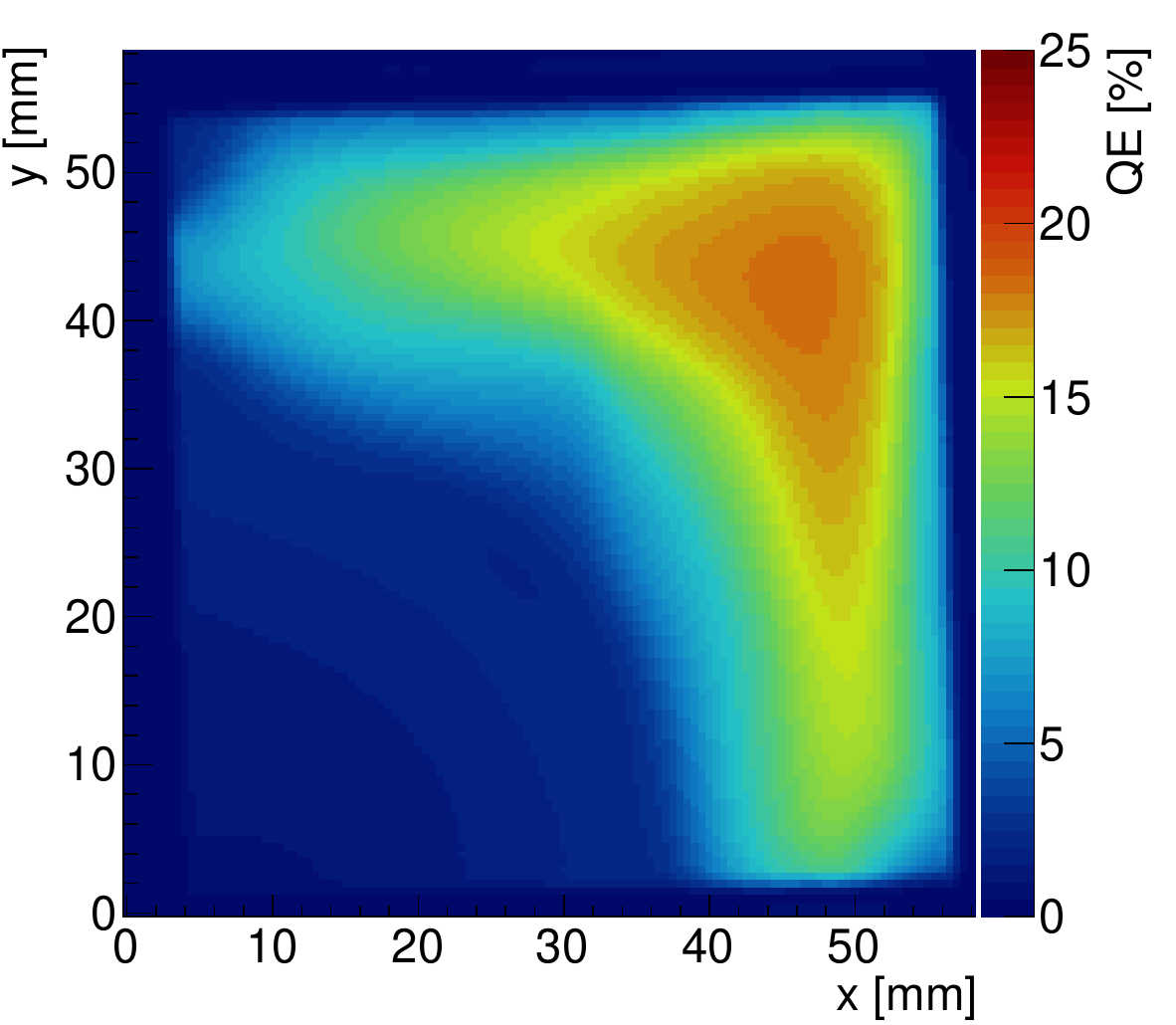}}
			\caption{QE scans of Photek A3191220 measured at a wavelength of \SI{372}{\nano \m} in \SI{0.5}{\mm} steps: (a) in February 2020 directly after the receipt of the tube, and (b) in March 2022 two years later.}
			\label{fig:qeleak}
		\end{figure}
		
		In fig. \ref{fig:qeleak} two QE surface scans of Photek A3191220 are exemplarily shown for different dates and collected IACs. Even though the left half of the sensor was covered and only the right side illuminated, the QE started dropping from the inital bad spot in the bottom left corner. A similar behavior was observed for the Photek A1200107 with the left half having been illuminated and the Photek A1200116 even without illumination at all \citep{daniel2022}. These results and the fact that all three MCP-PMTs have spots with low QE in some corners are strong indicators for vacuum microleaks. The problem seems to be solved meanwhile as the later produced Photek A2200606 does neither have evolving QE damages nor shows any QE loss up to \SI{2.4}{\coulomb \per \cm\squared}.
		

		\begin{figure}[ht]
			\centering
			\begin{subfigure}[ht]{0.49\linewidth}
				\includegraphics[width=\linewidth]{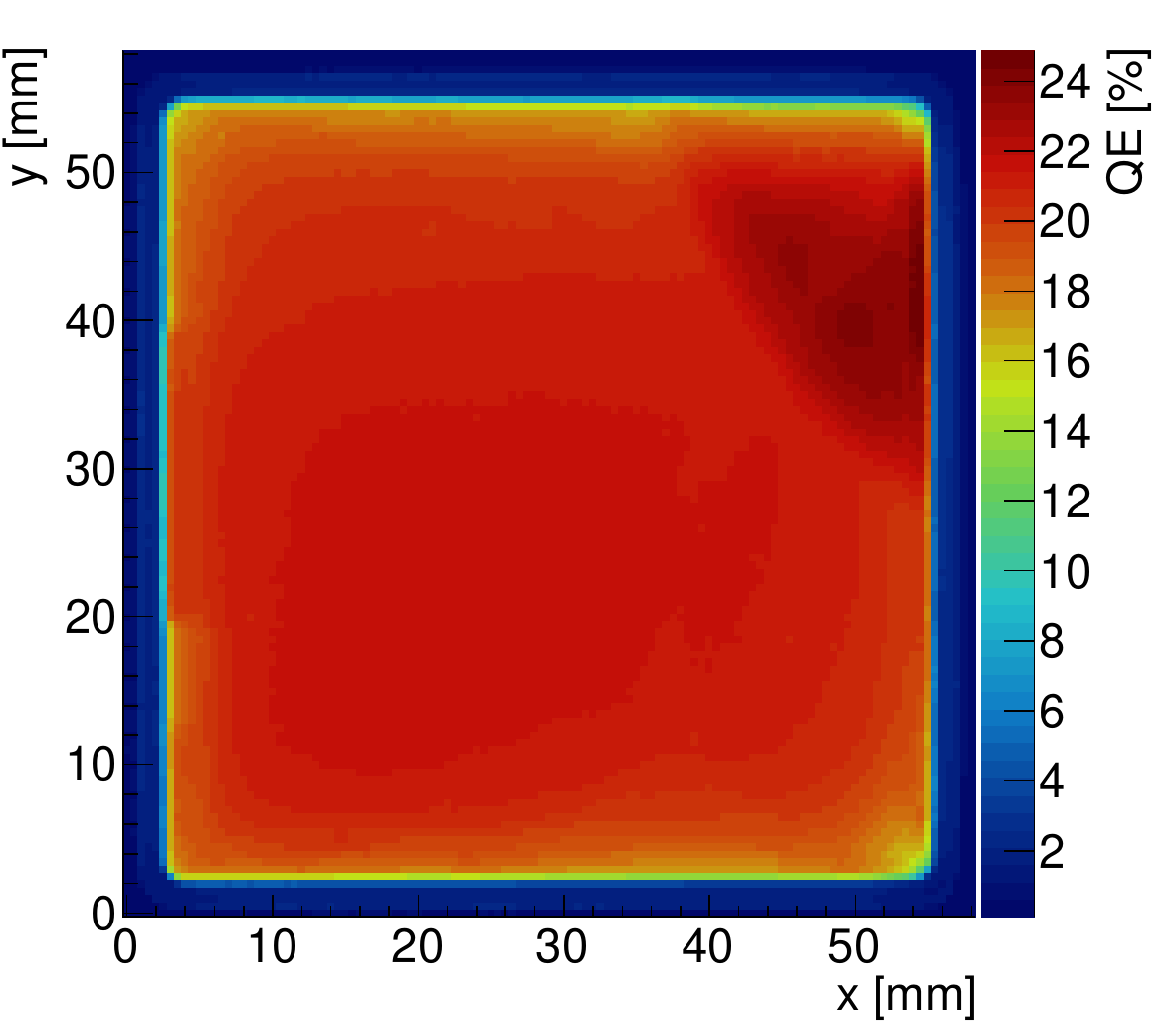} 
				\caption{QE scan Photonis 9002222.}
				\label{fig:qescan2222}
			\end{subfigure}
			\begin{subfigure}[ht]{0.49\linewidth}
				\includegraphics[width=\linewidth]{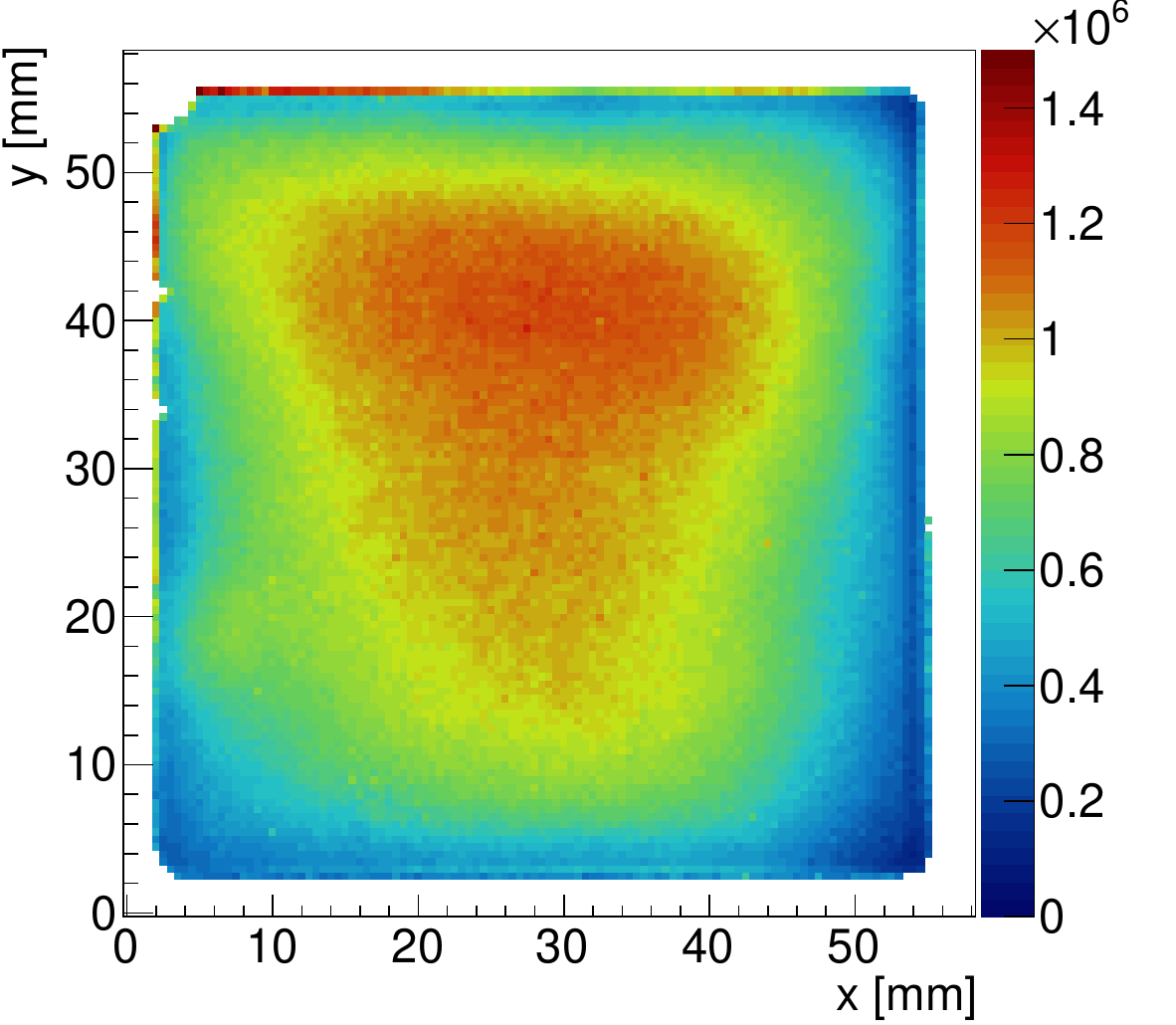} 
				\caption{Gain scan Photonis 9002223.}
				\label{fig:gainscan2223}
			\end{subfigure}
			\caption{QE (a) and gain (b) surface scans of two of the most recent Photonis MCP-PMTs measured at a wavelength of \SI{372}{\nano \m} in \SI{0.5}{\mm} step sizes.}
			\label{fig:surfacescans}
		\end{figure}
		
\begin{figure}[ht]
			\centering
			\includegraphics[width=0.9\linewidth]{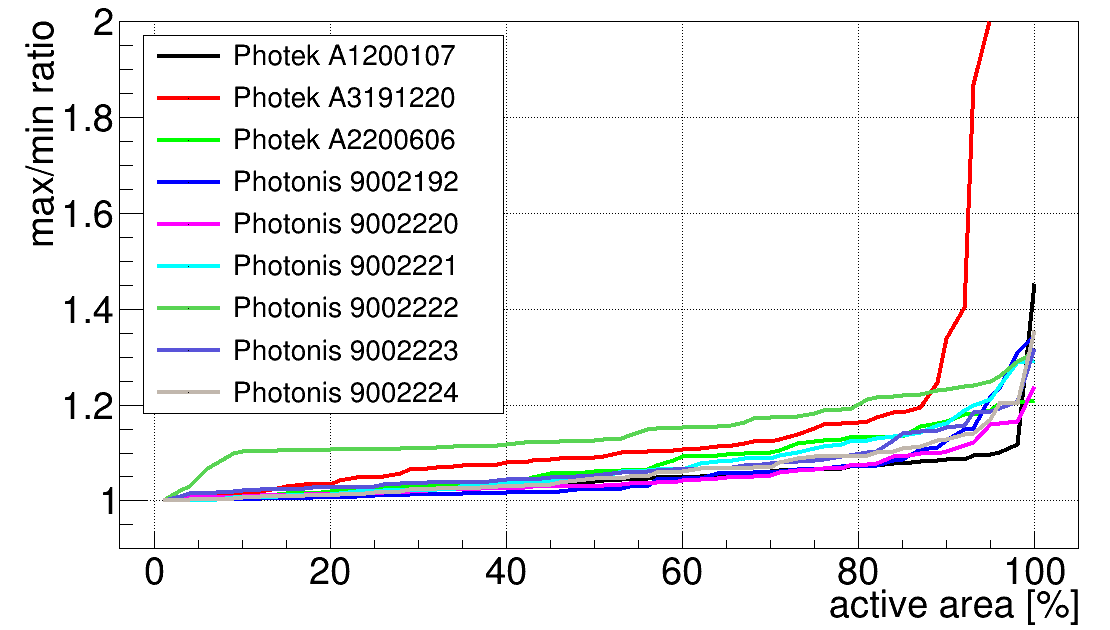} 
			\caption{Max/min ratio of the QE uniformity across the active area.}
			\label{fig:qe uniformity}
		\end{figure}

\begin{figure}[b]
			\centering
			\includegraphics[width=0.9\linewidth]{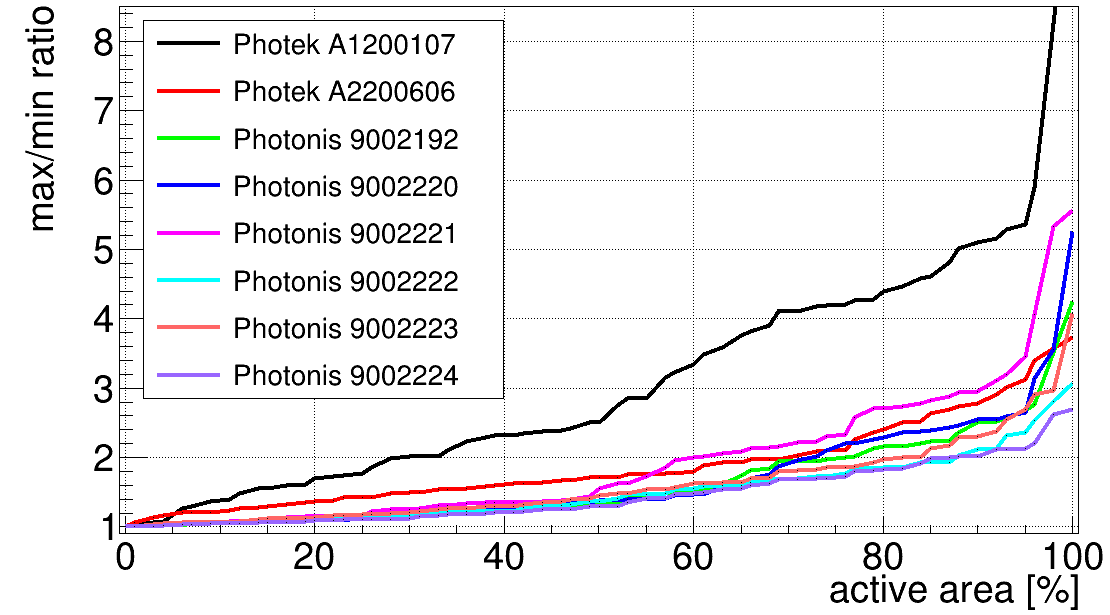} 
			\caption{Max/min ratio of the gain uniformity across the active area.}
			\label{fig:gain uniformity}
		\end{figure}

		In fig. \ref{fig:surfacescans} surface scans of two of the most recent Photonis MCP-PMTs are presented. Fig. \ref{fig:qescan2222} shows the QE spatial distribution of the Photonis 9002222 while in fig. \ref{fig:gainscan2223} the spatial distribution of the gain (corrected for QE) is presented. For an easier comparison of the quantitative values in terms of uniformity the mean QE and gain, respectively, are calculated for each anode pixel to determine a maximum to minimum ratio. Figures \ref{fig:qe uniformity} and \ref{fig:gain uniformity} show the QE and gain uniformity plotted versus the active area for the latest Photek and Photonis MCP-PMTs.
		
		For the PANDA Barrel DIRC a QE max/min ratio over the whole active area of less than 1.5 is required, which is met by all MCP-PMTs shown in fig. \ref{fig:qe uniformity} except for the Photek A3191220. It exceeds this limit due to the initial QE damage seen in fig. \ref{fig:qeA31}. For the gain uniformity a max/min ratio of less than 3 is required. This is fulfilled for at least \SI{90}{\%} of the active area for all sensors shown in fig. \ref{fig:gain uniformity} apart from Photek A1200107 which exceeds the required max/min ratio already after \SI{55}{\%} active area.

	\subsection{MCP-PMT characteristics inside magnetic fields} \label{ch:2.2}
	
\begin{figure}[ht]
			\centering
			\begin{subfigure}[ht]{0.32\linewidth}
				\includegraphics[width=\linewidth]{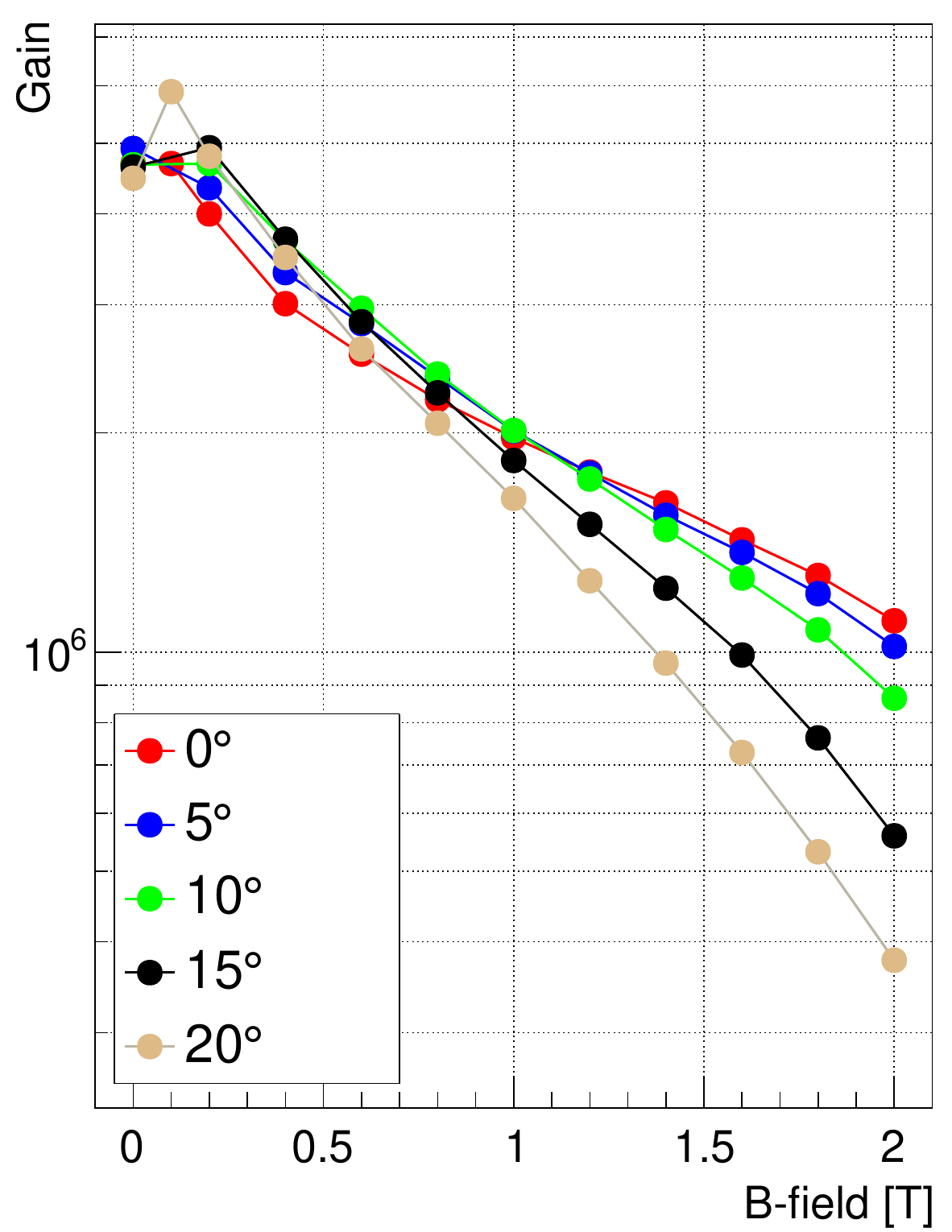} 
				\caption{Photek A1200116, \SI{2760}{\V} HV, \SI{6}{\micro\m} pore diameter, 8x8 anode pixels.}
				\label{fig:bfieldA12}
			\end{subfigure}
			\hfill
			\begin{subfigure}[ht]{0.32\linewidth}
				\includegraphics[width=\linewidth]{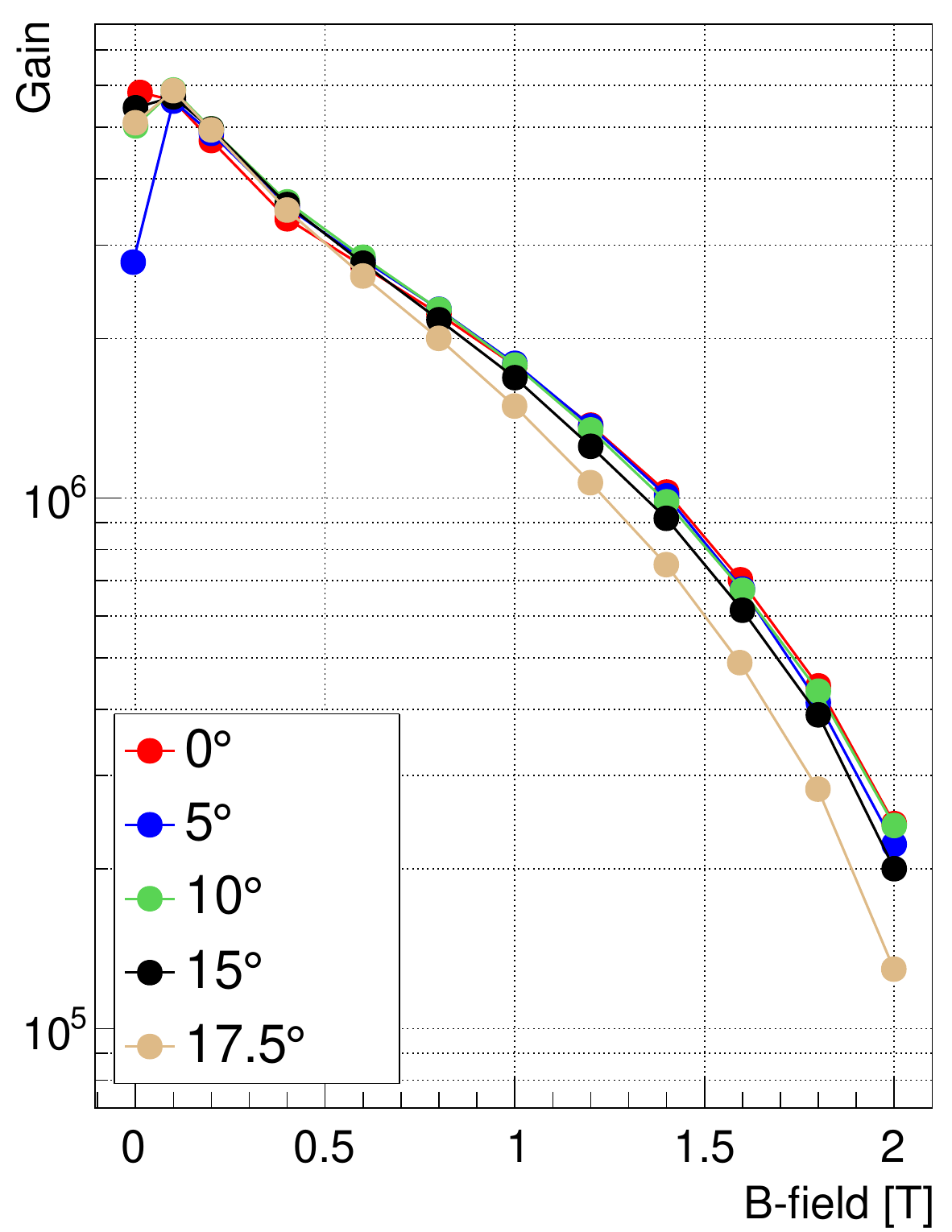} 
				\caption{Photonis 9002192, \SI{2700}{\V} HV, \SI{10}{\micro\m} pore diameter, 8x8 anode pixels.}
				\label{fig:bfield2192}
			\end{subfigure}
			\hfill
			\begin{subfigure}[ht]{0.32\linewidth}
				\includegraphics[width=\linewidth]{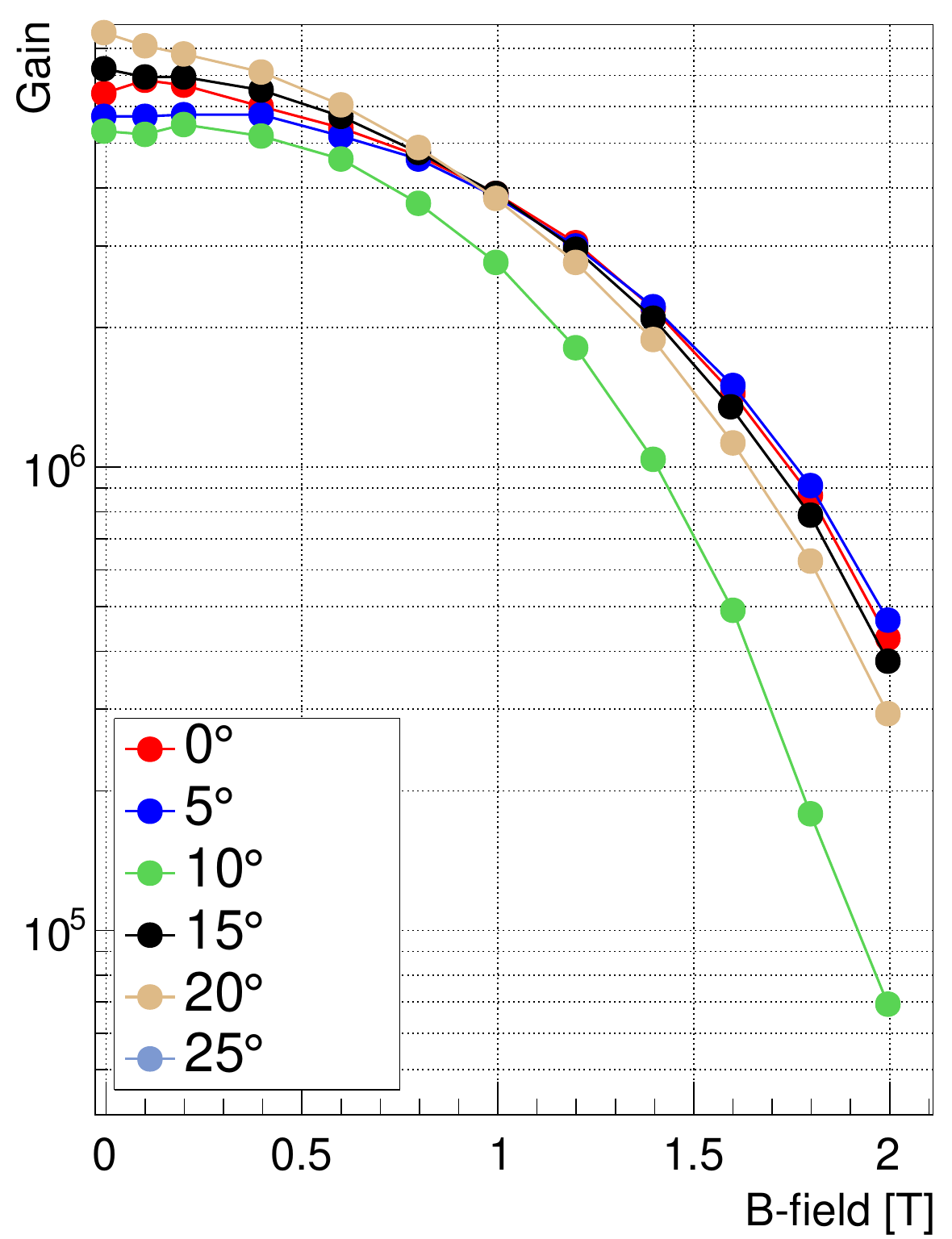} 
				\caption{Photonis 943P541, \SI{2950}{\V} HV, \SI{10}{\micro\m} pore diameter, 3x100 anode pixels.}
				\label{fig:bfield943}
			\end{subfigure}
			\caption{Gain versus magnetic field strength for different tilting angles between PMT-axis and field direction.}
			\label{fig:bfieldgains}
		\end{figure}

		Another important aspect of MCP-PMTs is their performance inside magnetic fields. Therefore several test measurements were performed using a dipole magnet with a field strength up to \SI{2.1}{\tesla}. In fig. \ref{fig:bfieldgains} the results of gain measurements with three different MCP-PMTs are shown where a central pixel was illuminated with single photon intensity for different field strengths and tilting angles.
		
		The gain behavior of Photek A12000116, Photonis 9002192 and 943P541 (3x100 anode pixels) is shown in fig. \ref{fig:bfieldgains} (\ref{fig:bfieldA12}, \ref{fig:bfield2192}, and \ref{fig:bfield943}, respectively). All tubes show different slopes of the gain behavior. The tubes with 8x8 anode pixels show a gain loss factor of \si{\sim 3} from \SIrange{0}{1}{\tesla} while the Photonis MCP-PMT with 3x100 anode pixels loses only a factor of 2 in gain due to a different internal geometry. For higher magnetic fields up to \SI{2}{\tesla} the Photonis MCP-PMTs show a significantly higher gain loss than the Photek tube. The main reason for the big difference in gain loss is the MCP pore diameter, \SI{6}{\micro\m} for the Photek MCP-PMT and \SI{10}{\micro\m} for the Photonis devices.
		
		\begin{figure}[ht]
			\centering
			\begin{subfigure}[ht]{0.49\linewidth}
				\includegraphics[width=\linewidth]{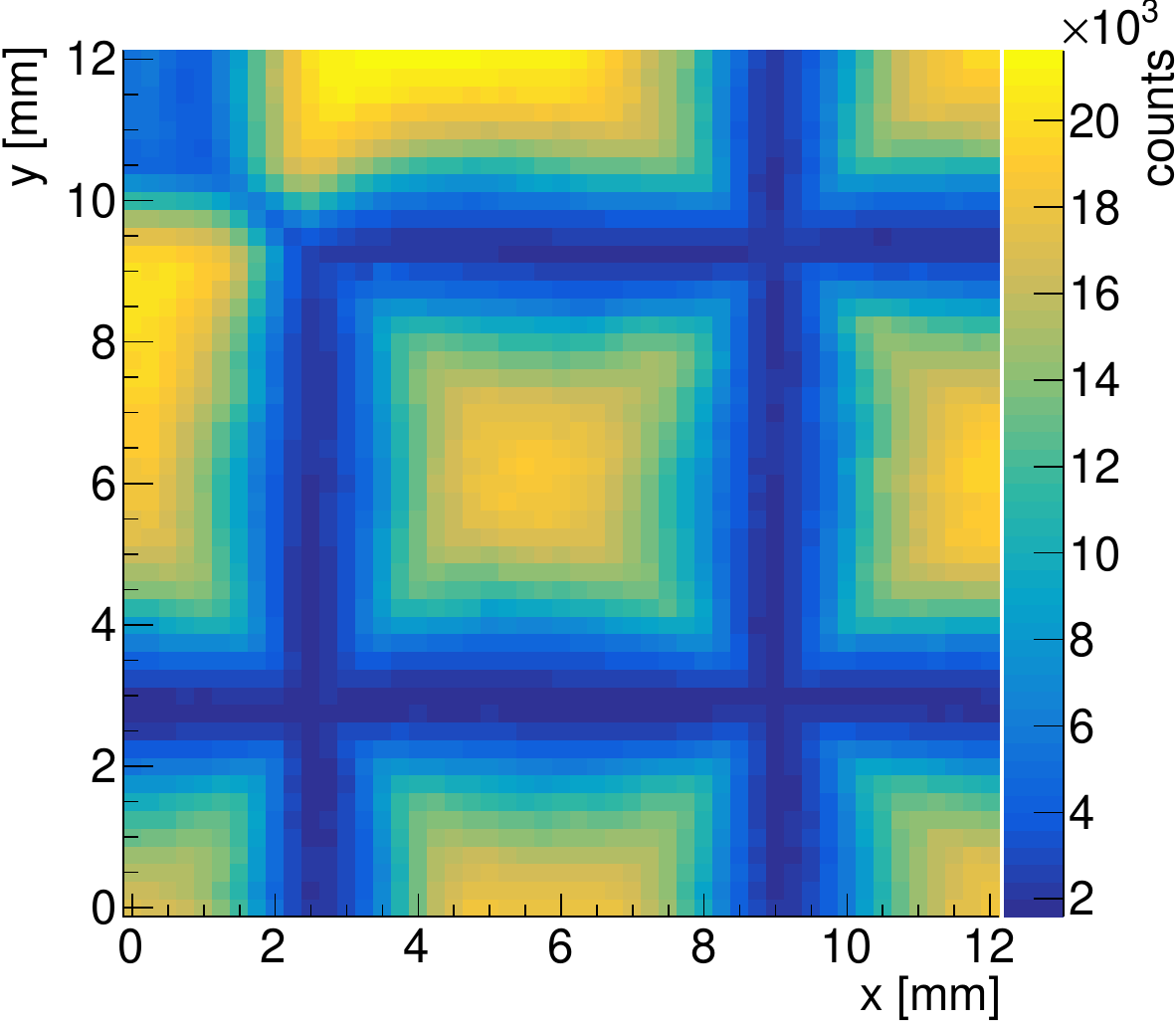} 
				\caption{Spatial distribution of one hit events at \SI{0.02}{T} field strength.}
				\label{fig:ct1}
			\end{subfigure}
			\begin{subfigure}[ht]{0.49\linewidth}
				\includegraphics[width=\linewidth]{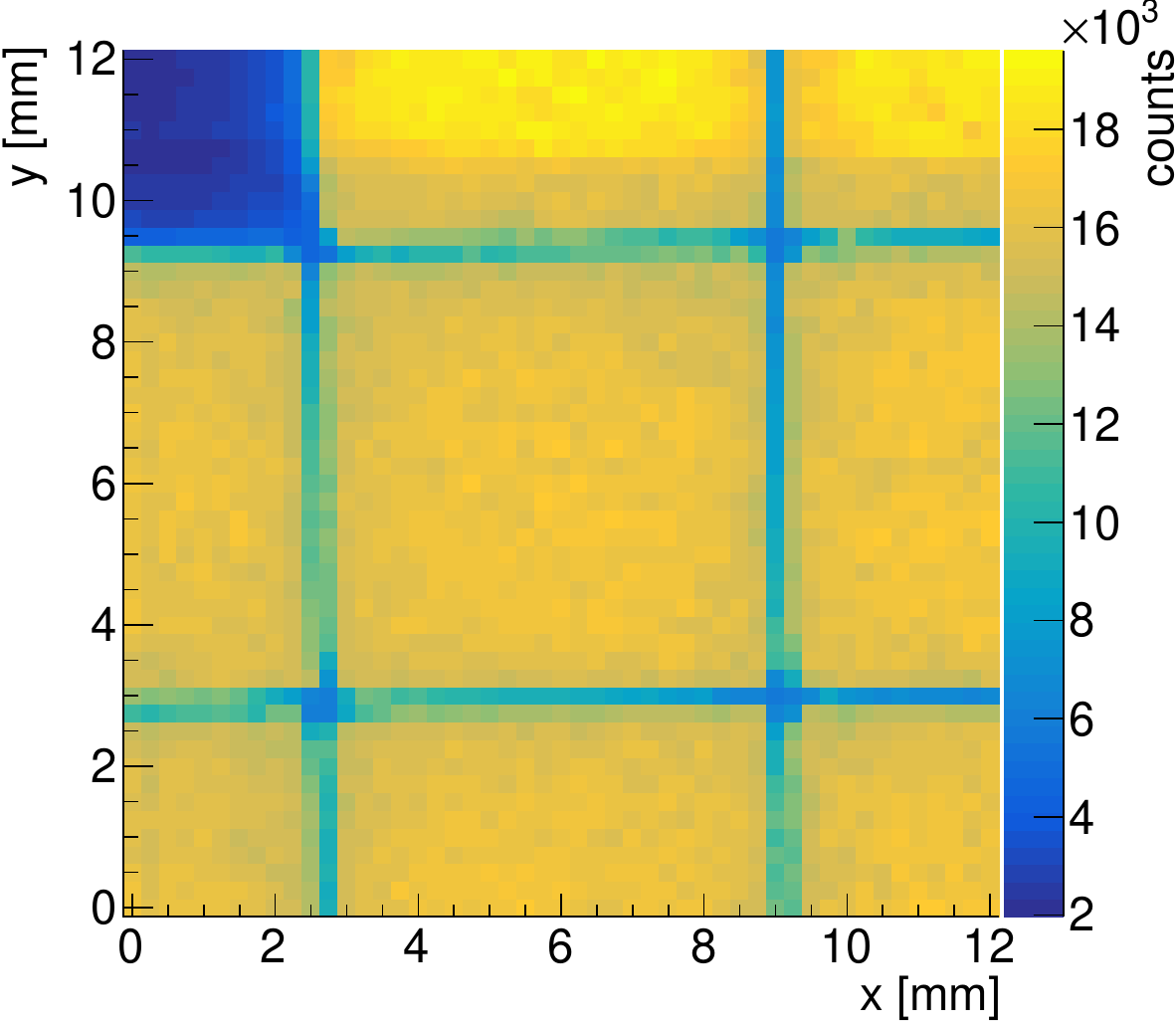} 
				\caption{Spatial distribution of one hit events at \SI{1.0}{T} field strength.}
				\label{fig:ct2}
			\end{subfigure}
			\caption{Spatial charge distribution for one hit events at a low and a high magnetic field strength. Only the inner 9 anode pixels of the MCP-PMT Photonis 9002192 were scanned. The upper left hole corresponds to a dead electronics channel.}
			\label{fig:crosstalk}
		\end{figure}

		Another investigated effect is the change of the charge cloud width with increasing magnetic field strength. In fig. \ref{fig:crosstalk} all events are plotted where only one of the 64 anodes detected a charge signal (= 1 hit). At \SI{0.02}{T} the pixels in fig. \ref{fig:ct1} are clearly separated since at this field the charge cloud at the anode is still quite wide and it is likely that the charge is spread across two or more pixels simultaneously. This produces two (or more) detected hits. At \SI{1.0}{T}, seen in fig. \ref{fig:ct2}, the gap between the pixels shrinks drastically because the charge cloud width and thus the probability of hitting two or more pixels at the same time is decreased. This shows that at a high magnetic field the spatial resolution of the PMT is improved.

	\subsection{Collection efficiency (CE) and time resolution} \label{ch:2.3}
		
		The CE describes the fraction of how many photoelectrons created at the PC lead to a measurable signal at the anode .
		
		The number of signals at the anode $\mathrm{N_{pe,~Anode}}$ can be extracted from a fit \citep{signalfit} to the charge distribution spectrum measured at a low laser frequency $\mathrm{f_{low}}$ (\SI{\sim 10}{\kilo \Hz}) in standard operation mode.
The number of photoelectrons $\mathrm{N_{pe,~PC}}$ created at the PC can only be measured indirectly. For this the sensor is illuminated at high frequencies $\mathrm{f_{high}}$ in the order of tens of MHz and operated in the QE setup mode (\SI{200}{\volt} between PC and MCP-in) where the current $\mathrm{I_{PC,f_{high}}}$ at the MCP-in is measured. In both setups the laser beam was split to additionally measure the laser intensity with a calibrated photodiode because the number of photons per pulse changes with frequency ($\mathrm{I_{diode,~f_{high}}}$, $\mathrm{I_{diode,~f_{low}}}$). The CE can then be calculated with the following formula:
		
		\begin{equation*}
			\mathrm{CE = \frac{N_{pe,~ Anode,~f_{low}} \cdot e \cdot f_{high}}{I_{PC,~f_{high}}} \cdot \frac{I_{diode,~f_{high}} \cdot f_{low}}{I_{diode,~f_{low}} \cdot f_{high}}}
		\end{equation*}
		
		\begin{table*}[ht]
		\centering
		\begin{tabular}{|c|c|c|c|c|c|c|c|} \hline 
			Tube & \makecell{Photonis \\ XP85112 \\ 9001394} & \makecell{Hamamatsu \\ R13266-07-64 \\ JS0022} & \makecell{Hamamatsu \\ R13266-07-64M \\ YH0250} & \makecell{Photonis \\ XP85112 \\ 9002108} & \makecell{Photek \\ MAPMT253 \\ A1200116} & \makecell{Photonis \\ XP85112 \\ 9002220}  \\ \hline 
			\makecell{Year of \\ production} & 2013 & 2014 & 2017 & 2018 & 2020 & 2022 \\ \hline
              & non-ALD & \makecell{ALD, film in \\ front of MCPs}  & ALD, no film & ALD, first Hi-CE & ALD & ALD, Hi-CE \\ \hline 
        		CE & \SI{63(6)}{\%} & \SI{39(4)}{\%} & \SI{65(7)}{\%} & \SI{95(9)}{\%} & \SI{90(9)}{\%} & \SI{95(5)}{\%} \\ \hline 
     	\end{tabular}
     	\caption{Collection efficiency measured for different types of MCP-PMTs.}
     	\label{table:CE}
		\end{table*}
		
		In table \ref{table:CE} the results of the CE for various MCP-PMTs from different production years and vendors are compared. Old sensors of Photonis like the 9001394 have a CE of \SI{\sim 65}{\%}. A comparable Hamamatsu MCP-PMT like the YH0250 has a similar CE. For older MCP-PMTs (JS0022) from Hamamatsu which had a PC protection film in front of the first MCP a significantly reduced CE was measured. Later Photonis has built the Hi-CE MCP-PMT 9002108 with an increased CE of up to \SI{95}{\%}. The newest MCP-PMTs of Photonis and Photek reach similar values.
		
		\begin{figure}[ht]
			\centering
			\begin{subfigure}[ht]{0.49\linewidth}
				\includegraphics[width=\linewidth]{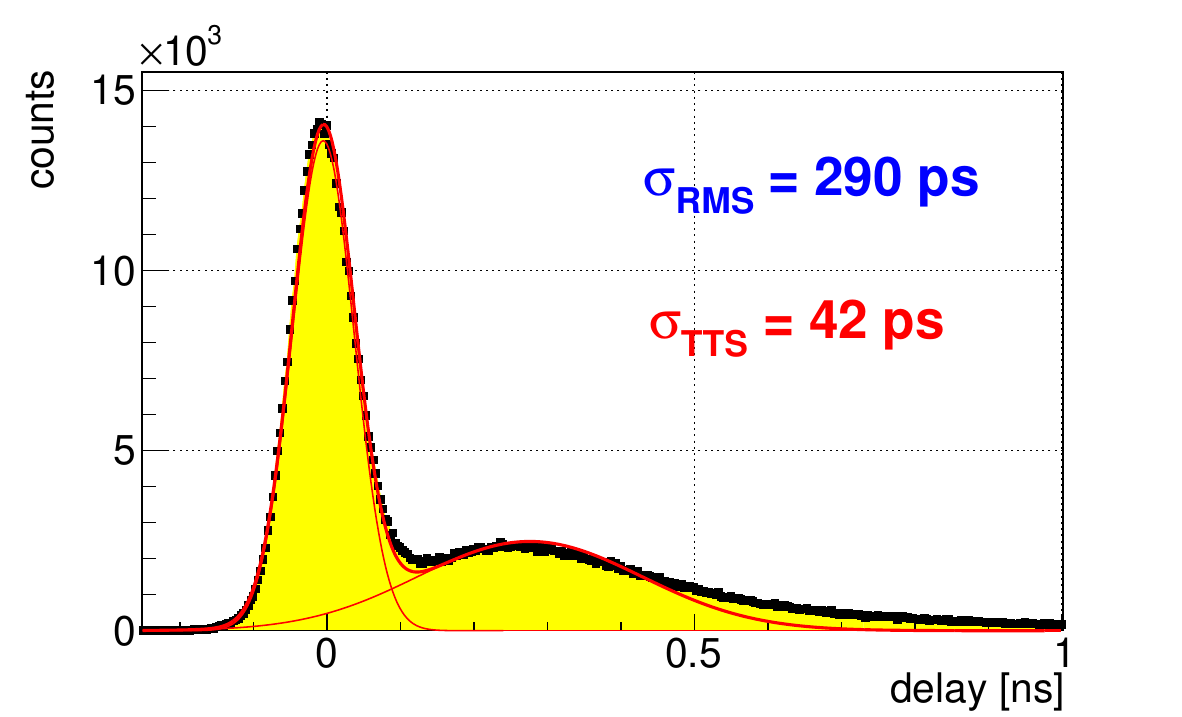} 
				\caption{Time resolution spectrum with \SI{\sim 200}{\volt} PC-MCP voltage.}
				\label{fig:timeres1}
			\end{subfigure}
			\begin{subfigure}[ht]{0.49\linewidth}
				\includegraphics[width=\linewidth]{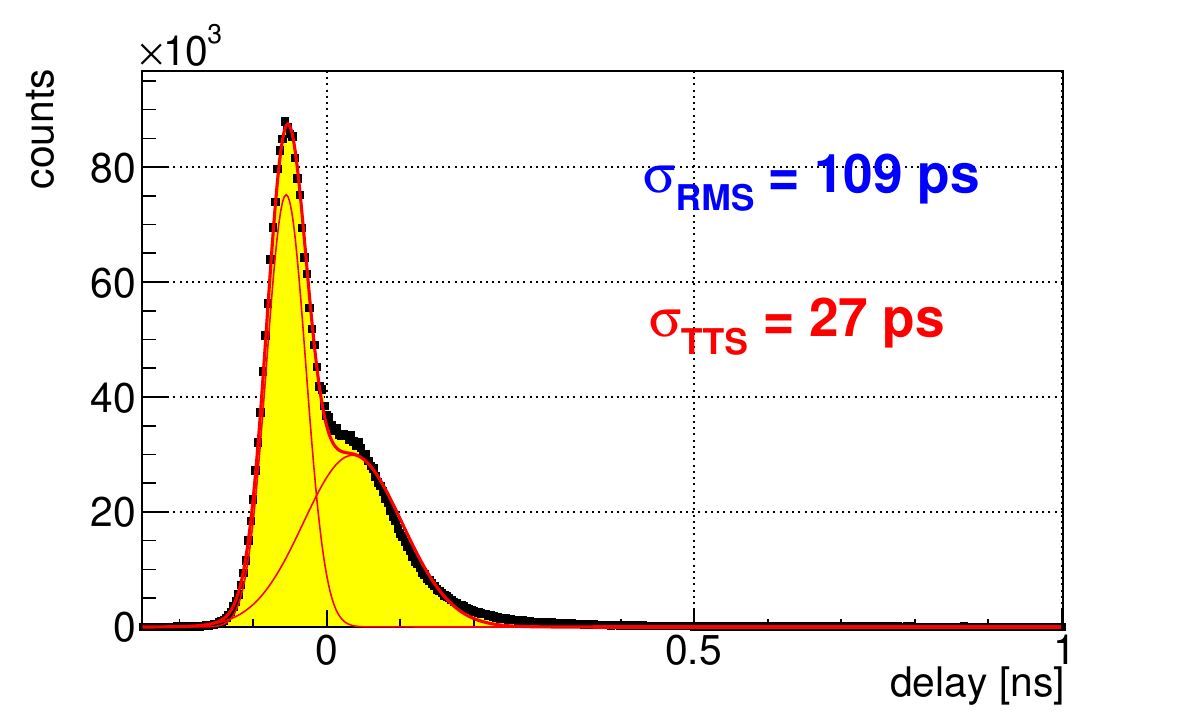} 
				\caption{Time resolution spectrum with \SI{\sim 700}{\volt} PC-MCP voltage.}
				\label{fig:timeres2}
			\end{subfigure}
			\caption{Time resolution of Photonis 9002193 at different PC-MCP voltages.}
			\label{fig:timeres}
		\end{figure}
		
		The increased CE was reached by capturing more electrons recoiling from the MCP-in and guiding them to the MCP pores for amplification. This however deteriorates the time resolution. In fig. \ref{fig:timeres1} the result of the time resolution measurement in standard operation setup (\SI{\sim 200}{\volt} PC-MCP voltage) is shown. One obtains a transit time spread $\mathrm{\sigma_{TTS}}$ of \SI{42}{\pico\s} and $\mathrm{\sigma_{RMS}}$ of \SI{290}{\pico\s} in the time window of \SIrange{-0.5}{2}{\nano\s}. By applying a higher voltage between PC and MCP-in (\SI{\sim 700}{\V}) the well separated recoil peak, seen in fig. \ref{fig:timeres1}, is now shifted into the main peak, shown in fig. \ref{fig:timeres2}. This leads to a significantly improved $\mathrm{\sigma_{RMS}}$ of \SI{109}{\pico\s}.
			
		\subsection{Rate capability} \label{ch:2.4}
		
		Due to the high interaction rates of up to \SI{20}{\mega \Hz} proton-antiproton annihilations in the PANDA experiment the MCP-PMTs have to be able to detect photons at high rates without gain loss. To obtain the results of the rate capability a reference diode as well as the whole MCP-PMT were illuminated homogeneously (by using a diffusor) and the current of the diode and that of all anode pixels shorted were measured. While the photon rate is increased the PMT anode current and that of the reference diode are compared. Once the shorted anode current increases at a lower rate than the diode current this is an indication of a gain saturation. The gain drop is caused by a too slow recharging of the MCP pores due to the high MCP resistance.
		
		\begin{figure}[ht]
			\centering
			\includegraphics[width=0.8\linewidth]{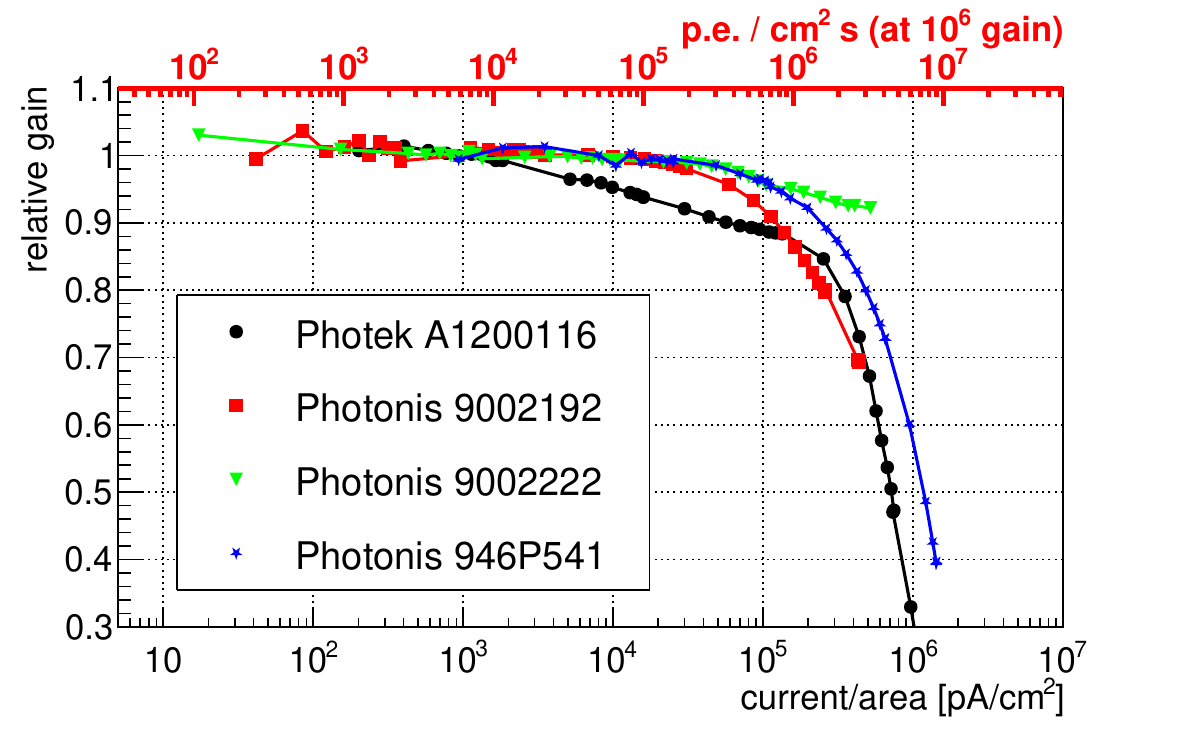} 
			\caption{Results of the rate capability measurement.}
			\label{fig:ratestab}
		\end{figure}	
		
		The results of these measurements are presented in fig. \ref{fig:ratestab} for different MCP-PMTs. For the PANDA Barrel DIRC a photon rate of a few \SI{100}{\kilo\Hz\per \cm\squared} is expected whereas it is \SI{\sim 1}{\mega \Hz \per \cm\squared} for the Endcap Disc DIRC. The sensors shown in fig. \ref{fig:ratestab} loose \SI{\sim 10}{\%} or less gain at \SI{1}{\mega \Hz \per \cm\squared} photon rate, so they are suitable for both DIRC detectors.

	\subsection{Results of measurements with the TRB/DiRICH system} \label{ch:2.5}
	
		The multihit capable TRB/DiRICH DAQ system \citep{Ugur_Traxler_12_TRB} is used to perform surface scans of the MCP-PMTs \citep{Albert_18_TRB}. By analysing the hit time and the time over threshold information for each hit of every anode channel and combining this with other measured quantities, e.g., the xy-position and the hit multiplicity one can deduce higher level internal PMT properties like afterpulse ratio, dark count rate, charge sharing and electronic crosstalk, and the time resolution.
		
		To deduce the afterpulse ratio the following approach is applied: in the analysis the same threshold is used for all hits and the measured hit time distribution (ranging from \SIrange{-10}{1}{\micro \s}) is shifted in such a way that the centroid of the  laser induced main peak is at \SI{100}{\nano \s}. The integral of this main peak is determined between \SI{99}{\nano \s} and \SI{104}{\nano \s}. The afterpulse time window is defined from \SIrange{104}{604}{\nano \s}. In addition, the number of dark count hits is calculated within a wide window before the main peak (e.g., between \SIrange{-1000}{0}{\nano \s}). Given that, the afterpulse ratio is simply defined as the ratio between the integrated hits within the afterpulse time window and the main peak integral. Both values are corrected for the number of darkcount hits appropriately scaled to the integration time windows.
		
		\begin{figure}[ht]
			\centering
			\begin{subfigure}[ht]{0.49\linewidth}
				\includegraphics[width=\linewidth]{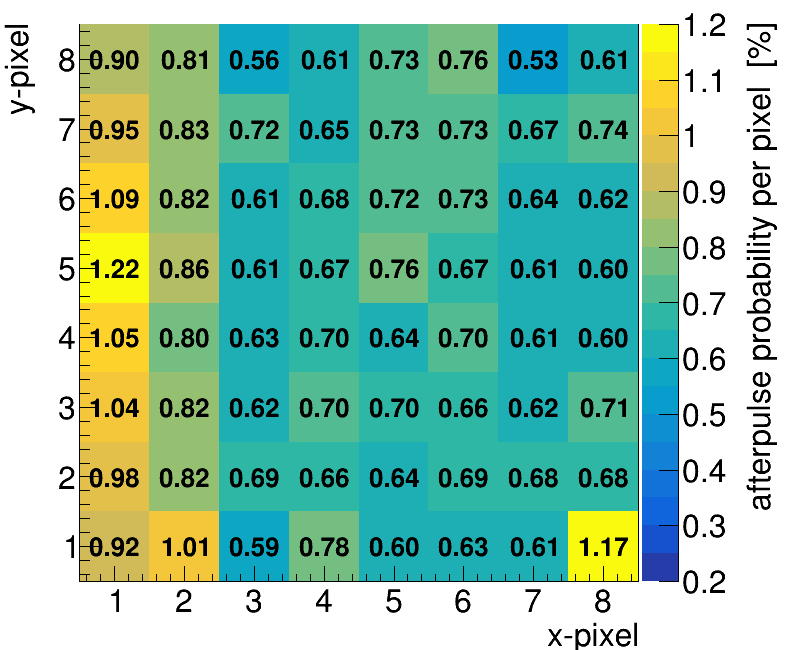} 
				\caption{Afterpulse probability per pixel.}
				\label{fig:ap}
			\end{subfigure}
			\begin{subfigure}[ht]{0.49\linewidth}
				\includegraphics[width=\linewidth]{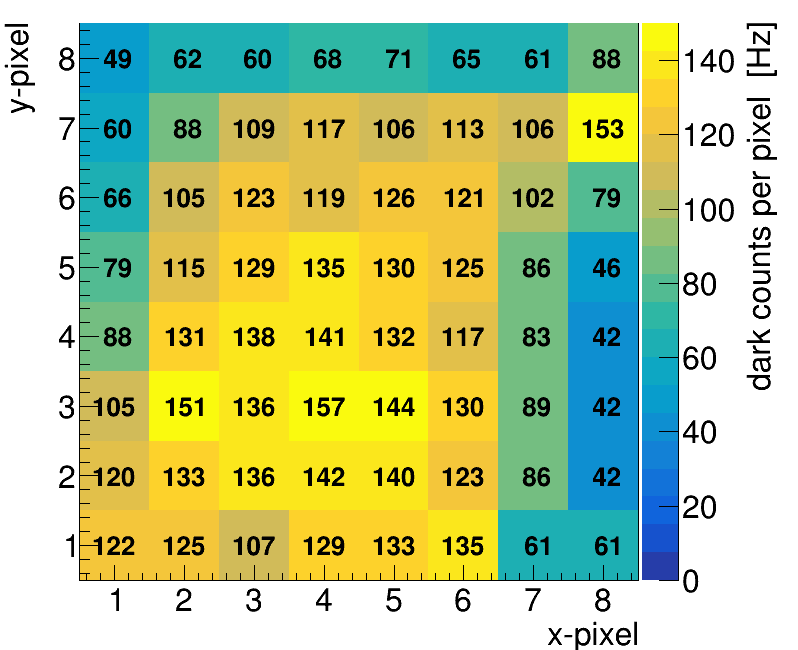} 
				\caption{Dark count rate per pixel.}
				\label{fig:dc}
			\end{subfigure}
			\caption{Afterpulse ratio and dark count rate of Photonis 9002223.}
			\label{fig:trb}
		\end{figure}
		
		In fig. \ref{fig:trb} the results of the afterpulse ratio (fig. \ref{fig:ap}) and dark count rate per pixel (fig. \ref{fig:dc}) are shown. A mean afterpulse ratio for the whole tube of \SI{0.74}{\%} and a mean dark count rate of \SI{104}{\Hz} per pixel corresponding to \SI{247}{\Hz\per\cm\squared} is received, which meet the requirements of the DIRC detectors.
	
	\subsection{Escalation} \label{ch:2.6}
		
		While performing test measurements with recent MCP-PMTs from Photonis an unexpected new behavior was observed. 
		When the MCP-PMT is operated at high gains ($\gtrsim \,$\num{5e6}) several not yet understood effects start to show up already at low (few \si{\kilo \Hz} single photons) or even without photon illumination. This also indicates that the effect is not correlated to the rate capability of the tube. At high photon rates the effect may already be triggered at lower gains. The main observations when escalation starts are:	
		
		\begin{itemize}
		\setlength\itemsep{0em}
			\item massive count rate increase
			\item significant gain drop
			\item drop of the MCP resistance
			\item significant production of white photons
		\end{itemize}

	
		
		\begin{figure}[ht]
			\centering
			\subcaptionbox{MCP and anode currents before and during escalation for the Photonis 9002222. The applied gain was \num{\sim 5e6} and there was no external photon illumination. \label{fig:esccurr}}[0.9\linewidth]{\includegraphics[width=\linewidth]{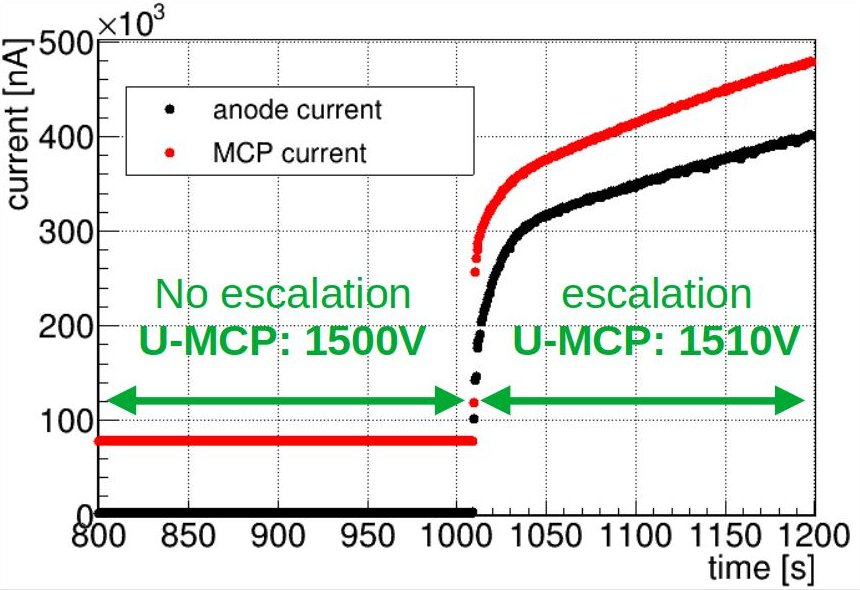}} %
			\subcaptionbox{Picture of the Photonis 9002193 built in measurement setup before operation. \label{fig:escref}}[0.4\linewidth]{\includegraphics[width=\linewidth]{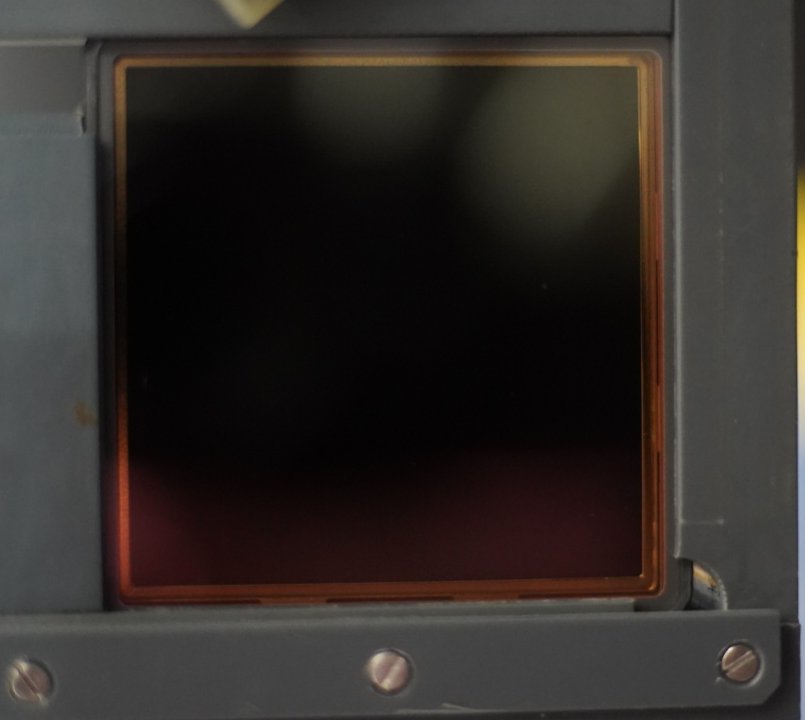}} %
			\subcaptionbox{Picture of the same MCP-PMT as in fig. \ref{fig:escref}, while operated in "escalation" mode without external illumination. \SI{20}{\s} exposure time of the photograph. \label{fig:escillu}}[0.4\linewidth]{\includegraphics[width=\linewidth]{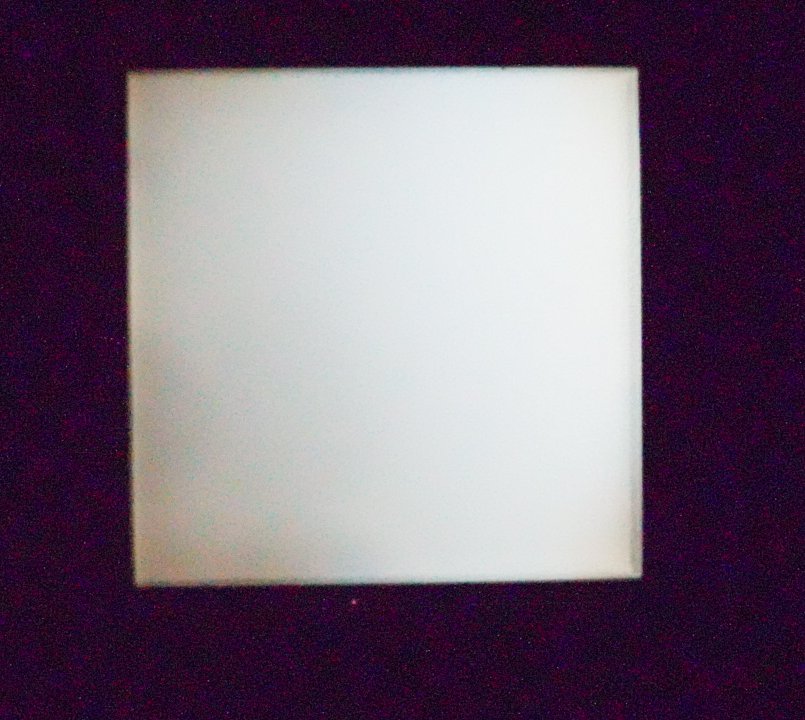}}
			\caption{Some escalation effects seen in the recent MCP-PMTs with two ALD layers.}
			\label{fig:escalation}
		\end{figure}
		
		In fig. \ref{fig:esccurr} the MCP and anode currents versus time are shown. A slight increase of the MCP voltage by \SI{10}{\volt} causes a massive increase of the currents at the anode and across the MCPs. In order to investigate the increased count rate, a digital camera was placed opposite to the MCP-PMT. In Fig. \ref{fig:escref} a reference picture of the tube before operation is displayed. Figure \ref{fig:escillu} shows the same MCP-PMT when operated in this "escalation" mode without ambient light and for a \SI{20}{\s} exposure time. One can clearly see that a massive number of photons is generated inside the sensor which explains the increased count rate. Further tests showed that these photons are created at or inside the MCP layers. At this point it is not yet clear if the observed escalation effect is of a similar origin as the signal induced noise reported in \citep{andreotti} and other references.

\section{Conclusion and outlook} \label{ch:kap3}
	
	Over the last years several significant improvements of MCP-PMTs were made. Especially the lifetime was increased enormously by applying an ALD technique to the pores. Furthermore, with the newest MCP-PMTs reaching \SI{\geq 90}{\%} CE, a  big improvement in detective quantum efficiency was accomplished. A somewhat worrisome observation is the "escalation" effect currently seen only in MCP-PMTs with two ALD layers. The consequences of this effect is being further investigated. Nevertheless, the recent MCP-PMTs are suitable for operation in both PANDA DIRC detectors. For the Barrel DIRC 155 Photonis MCP-PMTs of the type XP85112-S-BA were ordered. Delivery of the series production tubes to Erlangen started in May 2022 and seven units are currently undergoing detailed quality assurance measurements.


\bibliography{mybibfile}
 \bibliographystyle{elsarticle-num}





\end{document}